\documentclass[manuscript]{aastex631}
\usepackage{lineno}

\shorttitle{The September 30, 2020 event as observed by PSP$\slash$IS{\ensuremath{\odot}}IS}
\shortauthors{Getachew et al.}

\graphicspath{{./}{figures/}}

\begin{document}

\title{PSP$\slash$IS{\ensuremath{\odot}}IS Observation of a Solar Energetic Particle Event Associated With a Streamer Blowout Coronal Mass Ejection During Encounter 6}

\author[0000-0003-2408-4619]{T. Getachew}
\affiliation{Department of Astrophysical Sciences, Princeton University, Princeton, NJ 08544, USA}
\affiliation{Heliophysics Science Division, NASA Goddard Space Flight Center, Greenbelt, MD 20771, USA}
\affiliation{Physics Department, The Catholic University of America, Washington, DC 20064, USA}

\author[0000-0001-6160-1158]{D. J. McComas}
\affiliation{Department of Astrophysical Sciences, Princeton University, Princeton, NJ 08544, USA}

\author[0000-0002-3841-5020]{C. J. Joyce}
\affiliation{University of New Hampshire, Durham, NH 03824, USA}

\author[0000-0001-6590-3479]{E. Palmerio}
\affiliation{Space Sciences Laboratory, University of California--Berkeley, Berkeley, CA 94720, USA}
\affiliation{CPAESS, University Corporation for Atmospheric Research, Boulder, CO 80301, USA}

\author[0000-0003-2134-3937]{E. R. Christian}
\affiliation{Heliophysics Science Division, NASA Goddard Space Flight Center, Greenbelt, MD 20771, USA}

\author[0000-0002-0978-8127]{C. M. S. Cohen}
\affiliation{California Institute of Technology, Pasadena, CA 91125, USA}

\author[0000-0002-7318-6008]{M. I. Desai}
\affiliation{Southwest Research Institute, San Antonio, TX 78228, USA}
\affiliation{University of Texas at San Antonio, San Antonio, TX 78249, USA}

\author[0000-0002-0850-4233]{J. Giacalone}
\affiliation{Lunar \& Planetary Laboratory, University of Arizona, Tucson, AZ 85721, USA}

\author[0000-0002-5674-4936]{M. E. Hill}
\affiliation{Johns Hopkins University Applied Physics Laboratory, Laurel, MD 20723, USA}

\author[0000-0001-7224-6024]{W. H. Matthaeus}
\affiliation{University of Delaware, Newark, DE 19716, USA}

\author[0000-0002-4722-9166]{R. L. McNutt}
\affiliation{Heliophysics Science Division, NASA Goddard Space Flight Center, Greenbelt, MD 20771, USA}

\author[0000-0003-1960-2119]{D. G. Mitchell}
\affiliation{Johns Hopkins University Applied Physics Laboratory, Laurel, MD 20723, USA}

\author[0000-0003-4501-5452]{J. G. Mitchell}
\affiliation{Heliophysics Science Division, NASA Goddard Space Flight Center, Greenbelt, MD 20771, USA}
\affiliation{Department of Physics, George Washington University, Washington, DC 20052, USA}

\author[0000-0002-8111-1444]{J. S. Rankin}
\affiliation{Department of Astrophysical Sciences, Princeton University, Princeton, NJ 08544, USA}

\author[0000-0002-2270-0652]{E. C. Roelof}
\affiliation{Johns Hopkins University Applied Physics Laboratory, Laurel, MD 20723, USA}

\author[0000-0002-3737-9283]{N. A. Schwadron}
\affiliation{Department of Astrophysical Sciences, Princeton University, Princeton, NJ 08544, USA}
\affiliation{University of New Hampshire, Durham, NH 03824, USA}

\author[0000-0003-2685-9801]{J. R. Szalay}
\affiliation{Department of Astrophysical Sciences, Princeton University, Princeton, NJ 08544, USA}

\author[0000-0002-4642-6192]{G. P. Zank}
\affiliation{Department of Space Science, The University of Alabama in Huntsville, Huntsville, AL 35899, USA}

\author[0000-0002-4299-0490]{L.-L. Zhao}
\affiliation{Department of Space Science, The University of Alabama in Huntsville, Huntsville, AL 35899, USA}
\author[0000-0001-6886-855X]{B. J. Lynch}
\affiliation{Space Sciences Laboratory, University of California--Berkeley, Berkeley, CA 94720, USA}

\author[0000-0002-6924-9408]{T. D. Phan}
\affiliation{Space Sciences Laboratory, University of California--Berkeley, Berkeley, CA 94720, USA}

\author[0000-0002-1989-3596]{S. D. Bale}
\affiliation{Space Sciences Laboratory, University of California--Berkeley, Berkeley, CA 94720, USA}
\affiliation{Physics Department, University of California--Berkeley, Berkeley, CA 94720, USA}
\affiliation{The Blackett Laboratory, Imperial College London, SW7 2AZ London, UK}

\author[0000-0002-7287-5098]{P. L. Whittlesey}
\affiliation{Space Sciences Laboratory, University of California--Berkeley, Berkeley, CA 94720, USA}

\author[0000-0002-7077-930X]{J. C. Kasper}
\affiliation{BWX Technologies Inc., Washington, DC 20002, USA}
\affiliation{Department of Climate and Space Sciences and Engineering, University of Michigan, Ann Arbor, MI 48109, USA}

\newpage
\begin{abstract}

In this paper we examine a low-energy SEP event observed by IS{\ensuremath{\odot}}IS's Energetic Particle Instrument-Low (EPI-Lo) inside 0.18~AU on September 30, 2020. This small SEP event has a very interesting time profile and ion composition. Our results show that the maximum energy and peak in intensity is observed mainly along the open radial magnetic field. The event shows velocity dispersion, and strong particle anisotropies are observed throughout the event showing that more particles are streaming outward from the Sun. We do not see a shock in the in-situ plasma or magnetic field data throughout the event. Heavy ions, such as O and Fe were detected in addition to protons and 4He, but without significant enhancements in 3He or energetic electrons. Our analysis shows that this event is associated with a slow streamer-blowout coronal mass ejection (SBO-CME) and the signatures of this small CME event are consistent with those typical of larger CME events. The time--intensity profile of this event shows that PSP encountered the western flank of the SBO-CME. The anisotropic and dispersive nature of this event in a shockless local plasma give indications that these particles are most likely accelerated remotely near the Sun by a weak shock or compression wave ahead of the SBO-CME. This event may represent direct observations of the source of low-energy SEP seed particle population.
\end{abstract}

\newpage
\section{Introduction} \label{sec:intro}

Solar energetic particles (SEPs) in the heliosphere are accelerated from a few keV up to GeV energies by at least two mechanisms, namely reconnection (e.g., associated with solar flares) and coronal mass ejection (CME)-driven shocks. Particle populations associated with flares are known as impulsive SEP events, while particle populations accelerated by near-Sun CME-shocks are termed as gradual SEP events, and those associated with local CME-shocks are known as energetic storm particle (ESP) events \citep[for reviews, see e.g.,][]{Desai2016,Vainio2018}. The characteristics of impulsive and gradual SEP events at $\sim$1~AU can be found elsewhere \citep[for reviews, see e.g.,][]{Reames1999,Desai2016,Vainio2018} but, to summarize, impulsive events are usually less intense, more prompt and shorter-lived SEP fluxes than gradual events. Impulsive SEP events are typically electron rich, show enhancements in 3He/4He up to about 1000 times greater than coronal values, are associated with type III radio bursts, and have high charge states of heavy ions \citep[for reviews, see e.g.,][]{Reames1999,Mason2007,Desai2016,Vainio2018,Bucik2020}. Gradual SEP events are associated with CMEs and CME-driven shocks and are often accompanied by type II radio bursts. They are more intense than impulsive events and last longer, and their composition is similar to that of the solar corona.\\

The most extensive and detailed observations of SEPs have been made from spacecraft near 1~AU. Because of their low intensities and large observation distances (1~AU and beyond), the energetic particle environment of the quiet Sun, which is crucial for our understanding of the solar corona and solar wind, is not well understood. The quiet Sun is a solar region without sunspot-bearing active regions \citep{Bellot_Rubio2019} and includes solar features at granular \citep [smallest so far resolvable magnetic features by current instruments,][]{Ishikawa2008} and supergranular scales, which includes internetwork \citep{Livingston1975}, network field structures \citep{Sheeley1967}, and other small-scale active regions, such as coronal bright points \citep[CBPs;][]{Harvey1975}.\\

The connection between small-scale (quiet-Sun) and large-scale (active-region) magnetic structures as well as their generation mechanisms are topics of active research. Quiet-Sun magnetic field elements make a dominant contribution to the total magnetic field on the solar surface \citep{Getachew2019a,Getachew2019b,Mursula2021}, and may store and transfer huge amounts of energy to the upper atmosphere through different mechanisms. A CME can erupt from the quiet Sun, where the field is weak and no large filament or active region needs to be present in the pre-CME corona to initiate an eruption \citep[e.g.,][]{Robbrecht2009,Podladchikova2010,Vourlidas2018}. \\

Streamer-blowout coronal mass ejections (SBO-CMEs) are one of the manifestations of the quiet-Sun magnetic field. SBO-CMEs are usually slow (with an average speed of about 390\,km/s) and originate in the solar streamer belt. They are commonly characterized by a gradual swelling of the overlying streamer over a period of a few hours to a few days, followed by emergence of a bright and well-structured flux rope, and a generally slow CME from the streamer that leaves behind a depleted corona \citep{Illing1986,Sheeley1982}. Although SBO-CMEs show some variation with solar cycle, they do not follow the sunspot cycle, implying that they are not associated with active regions but originate in the quiet Sun. Their average duration, from the start of the streamer swelling to the release of the CME, is about 40 hours \citep{Vourlidas2018}. SBO-CMEs are observed only in streamer belts, and their locations follow the global dipolar field (the tilt of the global heliospheric current sheet). SBO-CMEs are typically stealth CMEs \citep[see][who first reported the eruption of a CME that left ``no trace behind'']{Robbrecht2009}, but in-situ signatures during their passage over a spacecraft do not differ significantly from those typical of interplanetary CMEs \cite[see][who analyzed the \citeauthor{Robbrecht2009} event at 1~AU]{Moestl2009,Lynch2010,Nieves-Chinchilla2011}. \citet{Lynch2016} suggested that SBO-CMEs are formed along the polarity inversion line below the streamer belt when magnetic energy accumulated by solar differential rotation is released via reconnection. \\

As the perihelion of the orbit of the Parker Solar Probe \citep[PSP;][]{Fox2016} spacecraft approaches closer and closer to the Sun (perihelion from 35\,solar radii ($R_{\odot}$) for
the first orbit to $<$10\,$R_{\odot}$ for the final three orbits), we are able to obtain in-situ measurements of solar output of plasma, energetic particles, and electromagnetic fields of the near-Sun environment that had not been previously explored \citep{Bale2019,Kasper2019,McComas2019,Howard2019}. The Integrated Science Investigation of the Sun \citep[IS{\ensuremath{\odot}}IS;][]{McComas2016} instrument suite, as part of the PSP mission, has observed several medium-sized SEP events, as well as weak, low-energy SEP events that are likely not detectable at 1~AU \citep{McComas2019}. We now have growing evidence that the existence of weak SEP events near the quiet Sun will enable a new way of interpreting SEP events associated with quiet-Sun magnetic structures \citep[e.g.,][]{McComas2019,Desai2020,Giacalone2020,Hill2020,Schwadron2020,Joyce2021a,Joyce2021b,Mitchell2020a,Mitchell2020b}. Recently, \citet{Joyce2021b} studied the radial evolution of energetic particles using PSP/IS$\odot$IS and Solar Terrestrial Relations Observatory Ahead \citep[STEREO-A;][]{Kaiser2008} spacecraft data, and showed that the properties of energetic particles observed at PSP's orbit and 1~AU are quite different, indicating that transport effects acted on the energetic particle populations and affected their properties in transit between the two spacecraft.\\

In this paper, we study the weak, low-energy SEP event of September 30, 2020 using PSP data from its sixth solar encounter around perihelion, and investigate the possible sources associated with this event. Our results show that the September 30, 2020 SEP event is weak, dispersive, and anisotropic and is associated with a slow SBO-CME observed by the coronagraph onboard STEREO-A off the solar east limb. The paper is organized as follows: Section~\ref{sec:remote} presents observations based on remote solar data. Section~\ref{sec:in-situ} focuses on the PSP$\slash$IS{\ensuremath{\odot}}IS in-situ observations. Finally, we discuss the results and present our conclusions in Section~\ref{sec:discussionandconclusion}.

\section{Remote Solar Observations} \label{sec:remote}

\begin{figure}[t]
	\centering
	\includegraphics[width=0.8\textwidth]{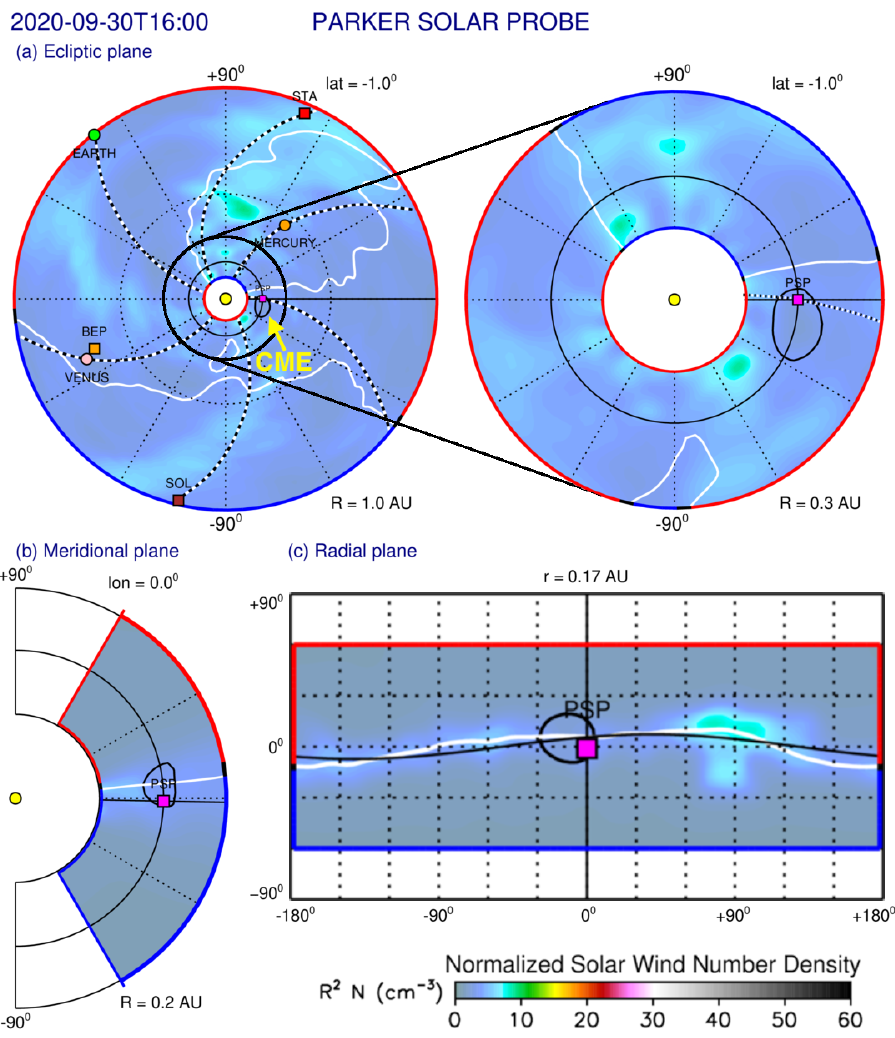}
	\caption{WSA--Enlil+Cone simulation results for the solar wind density on September 30, 2020 at 16:00\,UT.(a) Ecliptic plane, with the outer boundary set at (left) 1~AU and (right) 0.3~AU. (b) Meridional and (c) radial planes containing PSP. The radial location of PSP (magenta square) is shown as a thick black orbit and the CME (black contour) is indicated with a yellow arrow in panel (a). STEREO-A is marked by a red square. The simulation can be also found at NASA's Community Coordinated Modeling Center (CCMC), run id:  \href{https://ccmc.gsfc.nasa.gov/database_SH/Erika_Palmerio_072721_SH_1.php}{Erika\_Palmerio\_072721\_SH\_1}.}
	\label{fig:enlil_combo_plot}
\end{figure}

STEREO-A observed a slow and narrow CME on September 29, 2020 ejected from the eastern limb of the Sun, i.e., in close proximity to PSP's longitudinal location. Figure \ref{fig:enlil_combo_plot} shows the position of the planets and spacecraft within ${\sim}$1~AU from the Sun together with the weak CME as it passes PSP on September 30, 2020 at 16:00\,UT. The figure is generated using simulation results from the WSA--ENLIL+Cone model \citep{Odstrcil2003,Arge2004}. The different panels in the figure show results for the solar wind density in the ecliptic (out to 1~AU and zoomed-in to 0.3~AU), meridional, and radial planes. CME input parameters for the simulation were obtained via application of the Graduated Cylindrical Shell \citep[GCS;][]{Thernisien2011} model to simultaneous observations of the eruption in coronagraph data from the COR2 camera part of the Sun Earth Connection Coronal and Heliospheric Investigation \citep[SECCHI;][]{Howard2008} on board STEREO-A and the C3 camera part of the Large Angle Spectroscopic Coronagraph \citep[LASCO;][]{Brueckner1995} on board the Solar and Heliospheric Observatory \citep[SOHO;][located near Earth]{Domingo1995}. Figure \ref{fig:enlil_combo_plot} shows that PSP is magnetically connected to the CME, which grazes the spacecraft from its western flank. \\

\begin{figure}[t]
	\centering
	\includegraphics[width=0.8\textwidth]{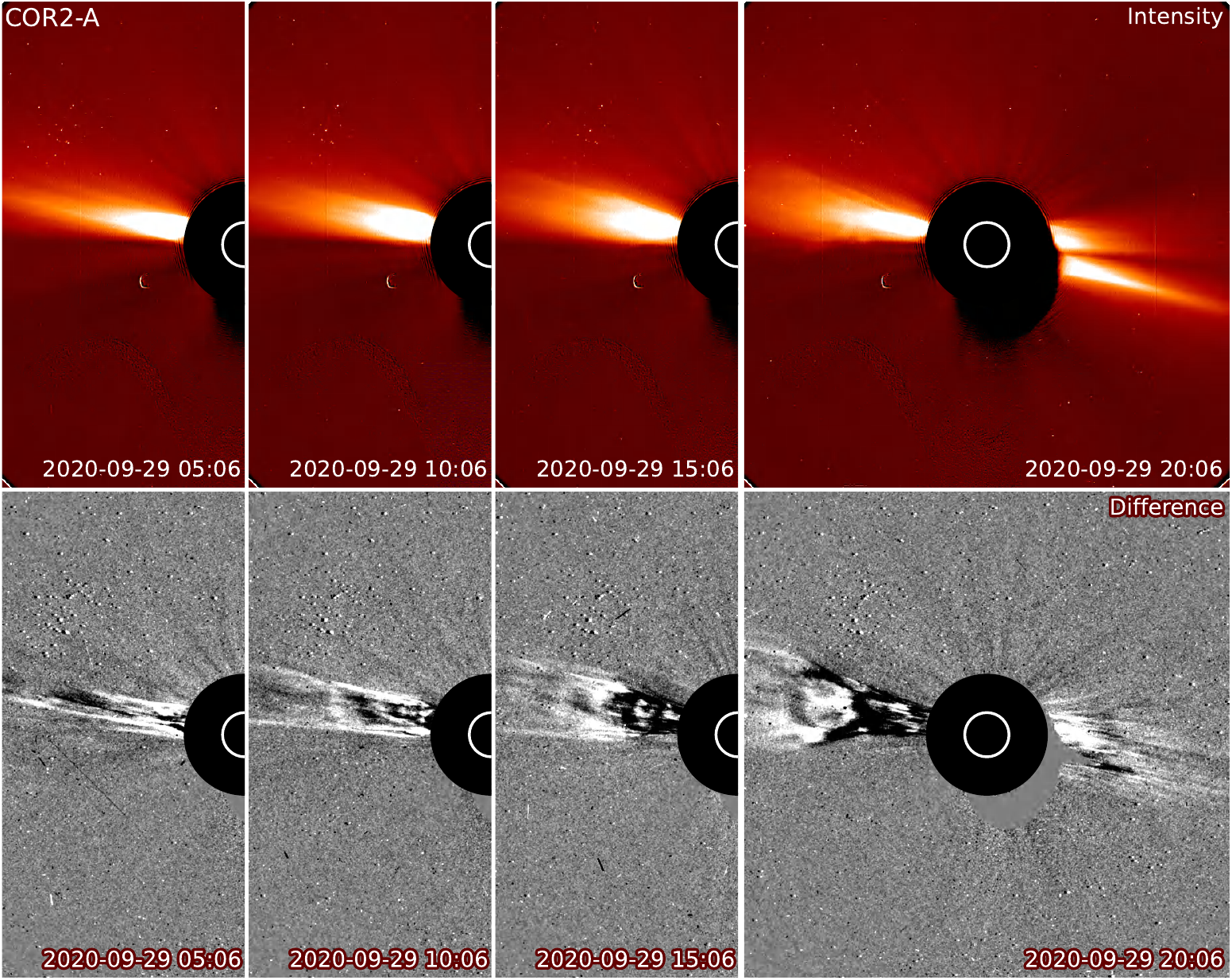}
	\caption{STEREO-A COR2 (upper panels) and COR2/RD (bottom panels) observation of the SBO-CME focused on the eastern limb on 29 September at 05:06, 10:06, 15:06 and 20:06 UT (from left to right). The animated version of this figure runs from September 28, 2020 at 12:00\,UT to September 30, 2020 at 12:00\,UT.\\ (An animation of this figure is available.)}
	\label{fig:cor2_figure}
\end{figure}

Figure \ref{fig:cor2_figure} shows STEREO-A COR2 (upper panels) and the corresponding running-difference (COR2/RD, bottom panels) images taken on September 29, 2020 at 05:06, 10:06, 15:06, and 20:06\,UT (from left to right). As can be seen from Figure \ref{fig:cor2_figure}, at the start of the event, a streamer off the solar eastern limb is seen to brighten and swell prior to the CME eruption (better seen in the associated movie). Subsequently, the CME is released slowly into the outer corona, followed by plasma outflows that also last for many hours. The CME leaves behind a depleted streamer, which is consistent with the properties of a typical streamer-blowout type CME \citep{Vourlidas2018,Korreck2020,Liewer2021}.\\

\begin{figure}[t]
	\centering
	\includegraphics[width=0.8\textwidth]{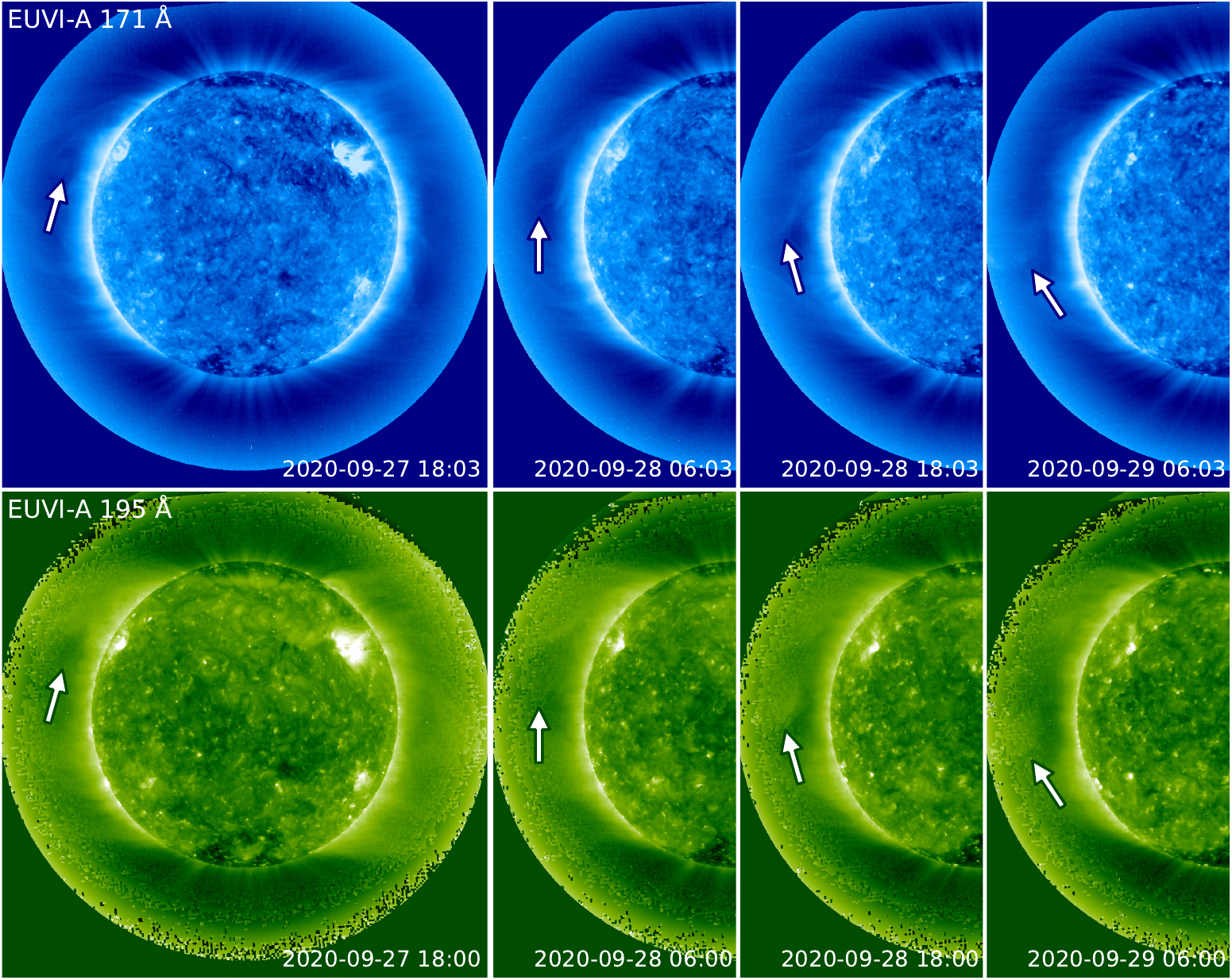}
	\caption{Remote-sensing observations of the September 29, 2020 SBO-CME by the EUVI telescope on board the STEREO-A spacecraft in the 171~{\AA} (upper panels) and 195~{\AA} (bottom panels) channels. The off-limb emission has been enhanced with a radial filter. The white arrows mark the core of the SBO-CME in the successive panels, showing its southward deflection. A flux rope-like structure (better seen in the animation) is observed on the northeastern limb on September 27, 2020 at about 18:00. The structure is channeled towards the solar equator through time. The animated version of this figure runs from September 27, 2020 at 12:00\,UT to September 30, 2020 at 00:00\,UT.\\ (An animation of this figure is available.)}
	\label{fig:euvi_figure}
\end{figure}

Figure~\ref{fig:euvi_figure} shows remote-sensing observations in extreme ultra-violet (EUV) of the September 29, 2020 SBO-CME taken by the SECCHI Extreme UltraViolet Imager (EUVI) instrument on board the STEREO-A spacecraft in the 171~{\AA} (upper panels) and 195~{\AA} (bottom panels) channels. As can be seen from Figure~\ref{fig:euvi_figure}, a flux rope-like structure (marked by the white arrows) slowly lifts off the northeastern limb (better seen in 171~{\AA}) starting on September 27, 2020 around 18:00\,UT, and is seen to deflect gradually towards the equator and the heliospheric current sheet. This structure moves in a ``rolling'' fashion (most evident in the associated movie), which is a common characteristics for slow, quiet-Sun eruptions during solar minimum \citep[e.g.,][]{Panasenco2013}. The CME started leaving the Sun already on late September 27, 2020, and completely left the EUVI field of view about 2~days later, which is a typical time-frame for SBO-CMEs \citep[e.g.,][]{Vourlidas2018,Liewer2021,Palmerio2021}. We note that the strong southward deflection observed in EUVI imagery is not reflected in COR2 data (Figure~\ref{fig:cor2_figure}), where the CME is seen to propagate almost radially along the direction of the overlying coronal streamer. This is consistent with previous findings, which showed that the most dramatic CME deflections and rotations tend to occur below ${\sim}5\,R_{\odot}$ due to magnetic forces in the low corona \citep[e.g.,][]{kay2015}.

\section{PSP/IS{\ensuremath{\odot}IS} In-situ Observations } \label{sec:in-situ}

\subsection{Orbit 6 Overview} \label{sec:orbit6}

IS{\ensuremath{\odot}IS} has been measuring solar energetic particles using the Energetic Particle Instrument-Low \citep[EPI-Lo,][]{Hill2017,Hill2020} and the Energetic Particle Instrument-High \citep[EPI-Hi,][]{Wiedenbeck2017}. EPI-Lo consists of eight wedges (W) with 10 time-of-flight apertures in each wedge (80 apertures in total) and provides observations of particles from 20 keV/nuc to 1.5 MeV/nuc over 2$\pi$ steradians. EPI-Hi consists of three telescopes of stacked solid-state detectors, including double and single-ended low-energy telescopes (LETs) and a double-ended high-energy telescope (HET). Collectively, EPI-Lo and EPI-Hi measure energetic particles of energies ranging from $\sim$0.2 to 200~MeV/nuc and species from protons to nickel, providing a comprehensive set of observations.\\

\begin{figure}[t]
	\centering
	\includegraphics[width=0.8\textwidth]{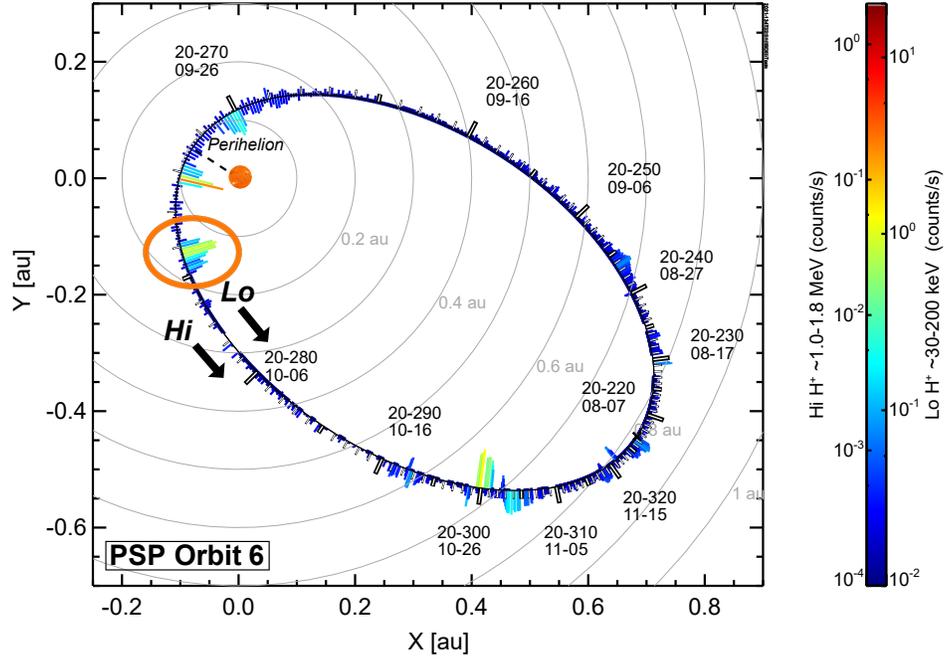}
	\caption{Overview of the geometry and the energetic particle measurements during orbit 6 of PSP following the plot format of \citet{McComas2019}. The EPI-Lo ion count rate summed over all apertures and energy ranges between 30--200 keV are shown inside the orbit, and the EPI-Hi count rate of LETA range 1 protons corresponding to about 1--2 MeV particles are shown outside the orbit. The SEP event analyzed in this paper (highlighted with an orange circle) is observed during PSP’s encounter 6 in the outbound portion of the orbit inside 0.18~AU. The Sun is shown as a small orange sphere (not to scale). The count rate levels are indicated by both the height and color of the bars. This event was detected by EPI-Lo on September 30, 2020 (DOY 274-2020) but not detected by EPI-Hi.}
	\label{fig:Orbit_06}
\end{figure}

PSP's orbit 6 spanned from August 2, 2020 (DOY 215-2020) through November 22, 2020 (DOY 327-2020). Encounter 6 (when PSP was inside 0.25~AU) started on September 21, 2020 and ended on October 2, 2020. Perihelion occurred on September 29, 2020 at a distance of 0.09~AU (19.35 $R_{\odot}$) from the center of the Sun. Figure~\ref{fig:Orbit_06} shows an overview of PSP’s orbit 6 geometry and the corresponding energetic particle measurements. The EPI-Hi LET1 A count rates (counts/sec) of protons and EPI-Lo count rates (counts/sec) of ions are shown on the outside and inside of the orbit, respectively. Color intensifications and taller bars indicate the occurrence of energetic particle events. The event analyzed in this paper is highlighted with an orange circle. As can be seen in Figure~\ref{fig:Orbit_06}, this event is detected by the EPI-Lo instrument, but is too small to extend into the EPI-Hi energy range $( \gtrsim 1 {\rm \; MeV} )$. The orbit 6 period is within the solar minimum phase and therefore populated with rather quiet energetic particle conditions. This gives a good opportunity to study the near-quiet-Sun energetic particle populations.

\subsection{The September 30, 2020 Event} \label{sec:Sep30}

\begin{figure}[t]
	\centering
	\includegraphics[width=\textwidth]{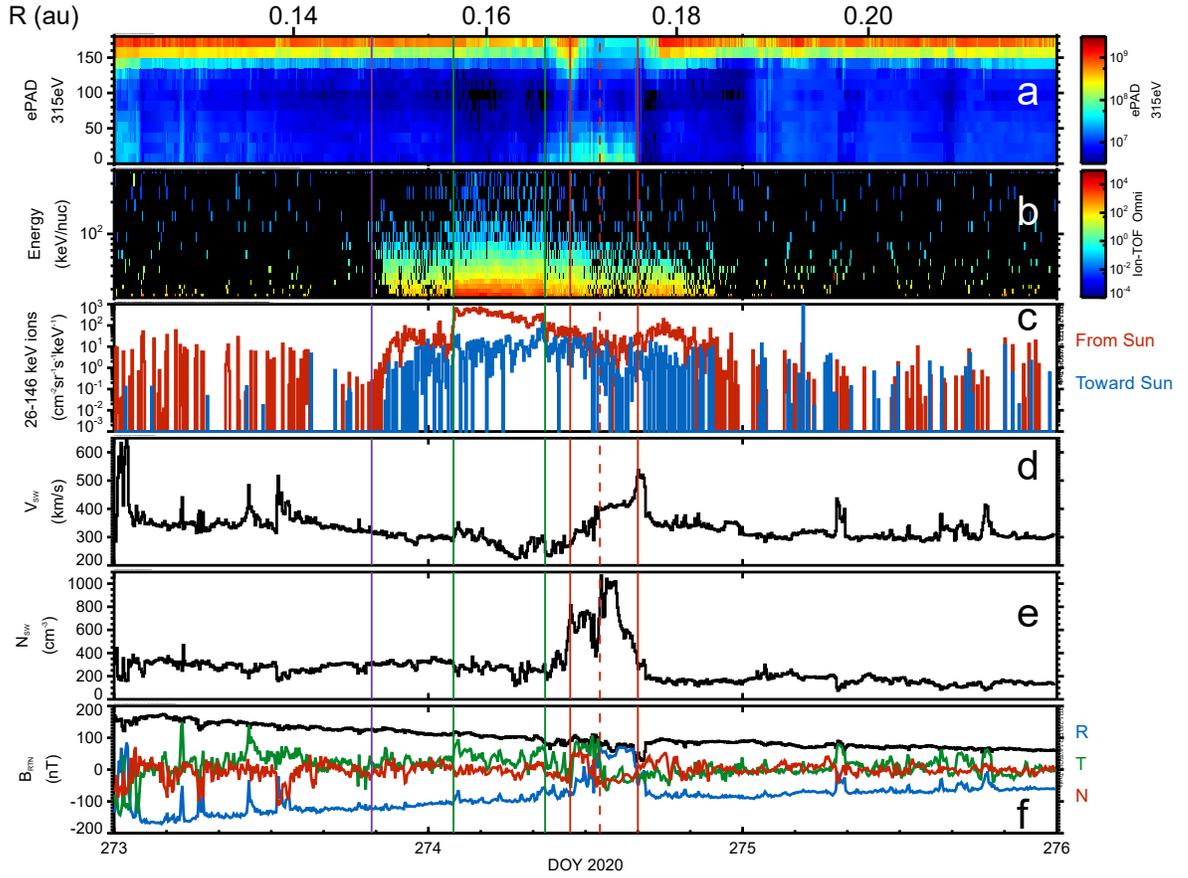}
	\caption{Overview plot of the CME Event that began on September 30, 2020. The top axis shows the spacecraft’s solar radial distance in AU. Panels from top to bottom are (a) 315 eV heat flux electron PADs from SPAN-E, (b) EPI-Lo spectrogram, (c) EPI-Lo intensities of ions (red and blue lines depict from and toward the Sun, respectively), (d) radial solar wind speed and (e) solar wind density measured by SPC and (f) magnetic field vector components (R-blue line, T-green line, and N-red line) and magnetic field strength (black line) as measured by FIELDS. All the data are averaged over 5-minute intervals. The SEP event onset is marked with a vertical purple line. The two vertical green lines represent the start and end time of the plateau (peak in intensity). The two vertical red lines bound the SBO-CME passage at PSP, and the dashed red line indicates the CME ejecta arrival time estimated by the WSA--Enlil simulation shown in Figure~\ref{fig:enlil_combo_plot}.}
	\label{fig:Figure_spe30_updated}
\end{figure}

\begin{figure}[t]
	\centering
	\includegraphics[width=\textwidth]{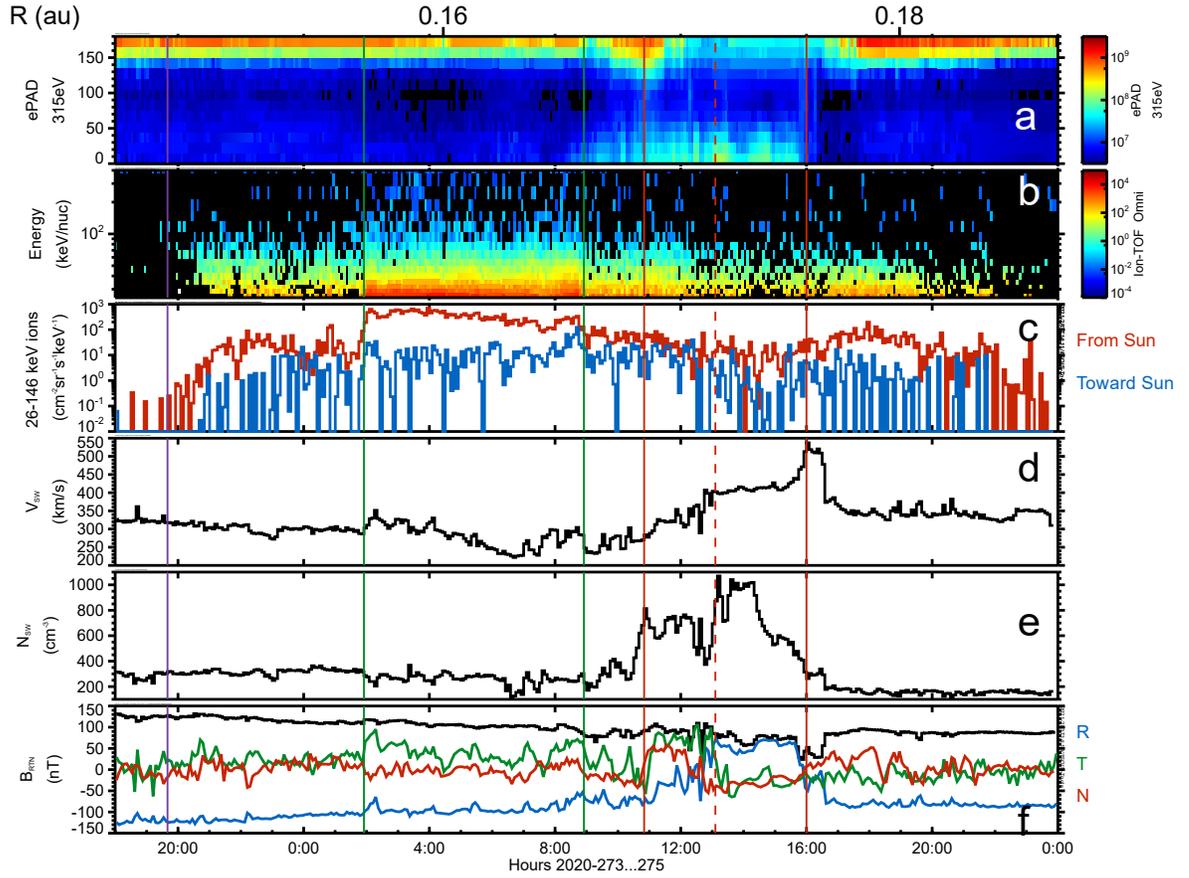}
	\caption{Zoomed-in version of Figure \ref{fig:Figure_spe30_updated}.}
	\label{fig:Figure_spe30_updated_zoomin}
\end{figure}

Figure \ref{fig:Figure_spe30_updated} shows an overview plot of the in-situ data for the September 30, 2020 SEP event as measured by PSP. Panel (a) shows the 315 eV heat flux strahl electron pitch angle distribution (PAD) from the Solar Probe Analyzer--Electron \citep[SPAN-E;][]{Whittlesey2020} instrument part of the Solar Wind Electrons Alphas and Protons \citep[SWEAP;][]{Kasper2016} investigation. Panel (b) shows the spectrogram for time-of-flight (TOF) only ions from IS{\ensuremath{\odot}IS}$\slash$EPI-Lo. The spectrogram is obtained by averaging over all the apertures except for 25, 31, 34, 35, and 44, which have a high rate of photon-induced accidentals due to punctures in its foil window from dust particle collisions \citep{Hill2020,Szalay2020}. We also calculated the intensity of the energetic particles moving away from and toward the Sun. The intensity of particles moving away from the Sun is obtained from the Ion-ToF particle intensities measured through the apertures of the EPI-Lo wedges looking in the sunward direction (W2, W3, and W4), while the intensity of particles moving toward the Sun is obtained from the Ion-ToF particle intensities measured through the apertures of the wedges looking in the anti-sunward direction (W0, W7, and W6). Panel (c) depicts the intensity of energetic particles moving from and toward the Sun obtained from IS{\ensuremath{\odot}IS}$\slash$EPI-Lo, demonstrating that this event was anisotropic. Panels (d) and (e) show the solar wind density and radial speed, respectivly, at PSP from the Solar Probe Cup \citep[SPC;][]{Case2020} part of SWEAP. Panel (f) shows the magnetic field vectors from the FIELDS instrument \citep{Bale2016} in the radial--tangential--normal (RTN) coordinate system. The position of PSP (in au) relative to the solar center is shown at the top of Figure \ref{fig:Figure_spe30_updated}. Figure \ref{fig:Figure_spe30_updated_zoomin} is a zoomed-in version of Figure \ref{fig:Figure_spe30_updated}, for better visibility of the event. \\

As can be seen from the ion spectrogram (panel (b) of Figure \ref{fig:Figure_spe30_updated}), the SEP event onsets at about 19:40\,UT on September 29, 2020 (vertical purple line) and shows clear velocity dispersion signatures, with the fastest particles arriving first. Similar dispersive events for low-energy ions associated with slow CMEs have been observed by the EPI-Lo instrument. The events observed on November 5, 2018 \citep{Hill2020}, November 11, 2018 \citep{Giacalone2020,Mitchell2020a}, and January 20, 2020 \citep{Joyce2021b} are some examples of weak SEP events that show clear velocity dispersion. A dispersive event indicates that the source of the energetic particles is remote, instead of being local. \\

Figure \ref{fig:Figure_spe30_updated} shows that the energy of ions starts to increase up to a couple hundred keV at about 02:00 on September 30, 2020 (DOY 274-2020) when solar wind speed, density, and magnetic field (particularly in the T-component) properties start to change, indicating the passage of a new field structure. The strongest intensity for this event is observed in the time interval between 02:00 and 08:50\,UT on September 30 (bounded by two green vertical lines) and shows a flat (constant) intensity--time profile throughout the interval, which is slightly different from the rest of the event. The flat intensity--time profile that persisted for about 8\,hours suggests a constant acceleration of particles of that energy \citep{Reames1999} and prolonged magnetic connectivity. The plasma and magnetic field data show that this enhancement (peak) in energetic particles seem to be associated with extremely slow solar wind speed and open magnetic field structure (indicated by the unidirectional strahl electron flow) and that the enhancement is observed before the predicted arrival time of the CME (dashed vertical red line). As shown in panel (a) of Figure \ref{fig:Figure_spe30_updated}, the strahl electron flow is unidirectional throughout the event except in the time interval between 10:00 and 16:00\,UT on September 30, 2020 (DOY 274-2020), during which the flow is bidirectional. A unidirectional electron flow implies the magnetic field lines are open or attached to the Sun only from one end, whereas a bidirectional electron flow represents a closed magnetic field structure in which the field lines are attached to the Sun at both ends \citep{Gosling1987}.    \\

A dropout in energetic particles is observed in the time interval between 11:00 and 16:00\,UT. The dropout is clearly seen in both the spectrogram and intensity plots shown in panels (b) and (c) of Figure \ref{fig:Figure_spe30_updated}. This dropout coincides with the CME arrival time as predicted by WSA--Enlil (vertical dashed red line), a bidirectional flow of strahl electrons, and a somewhat smooth rotation of the magnetic field vectors. The solar wind speed features an increasing profile and the plasma density is rather high, indicating a structure that is being compressed from behind by a faster solar wind flow \citep{Pizzo1978,Gosling1996}. Note that a bidirectional flow of strahl electrons \citep{Gosling1987} and a smooth rotation of the magnetic field components \citep{Burlaga1981} are commonly used as indicators of a CME magnetic structure, which we identify to encounter PSP between ${\sim}$11 and 16\,UT on September 30, 2020 (solid vertical red lines in Figure~\ref{fig:Figure_spe30_updated}). We note that the magnetic field strength (panel (f)) of the identified CME is not significantly higher than its surroundings, as is common in CME ejecta \citep[e.g.,][]{Kilpua2017}. However, we consider that (1) the weak, slow CME observed in remote-sensing imagery likely featured no strong, intrinsic magnetic fields, and (2) the trend of the magnetic field components ($B_{T}$ and $B_{N}$ rotate in the first half of the ejecta and stay around zero during the second half, while $B_{R}$ rotates from negative to positive) are suggestive of a flank/leg encounter, in agreement with the WSA--Enlil simulation results (Figure~\ref{fig:enlil_combo_plot}). The ion dropout during the CME arrival time is not surprising since energetic particles are often suppressed within CME magnetic clouds \citep{Forbush1937,Cane2000}.

We do not find signatures of a local shock in the plasma or magnetic field data shown in Figures \ref{fig:Figure_spe30_updated} and \ref{fig:Figure_spe30_updated_zoomin}. The absence of a local shock may indicate that these particles were remotely accelerated somewhere near the Sun. 150\,kev/nuc protons can travel the spiral-distance between the solar low corona and PSP ($\sim$0.14~AU) in shortly over a hour, hence they would have departed around 18:30\,UT on September 30, 2020, when the CME was already well into the outer corona (see its location in the corona a day earlier in Figure~\ref{fig:cor2_figure}). As discussed in Section \ref{sec:remote}, PSP was well connected to the CME since it encountered the eruption in situ, but such encounter was determined to be a flank one through both WAS--Enlil simulation results (Figure \ref{fig:enlil_combo_plot}) and in-situ observations (Figure~\ref{fig:Figure_spe30_updated_zoomin}). Hence, it is possible that particles traveled toward PSP once the western flank of the CME became magnetically connected to it as a consequence of expansion. Hence, it is possible that the CME initially drove a (weak) shock near the Sun. The profile of the SEP event studied here is consistent with that of magnetic connectivity achieved from the west of the observer \citep[see Figure 3.4 in ][]{Reames1999}, in agreement with our interpretation mentioned above. In this case, the strongest acceleration is assumed to occur near the central region of the shock, where the shock is presumably strongest and the speed is likely to be highest, and tends to decline around on the flanks \citep[see, e.g.,][]{Reames1999}. It is worth noting that the strongest acceleration typically reflects the geometry of the shock and magnetic field, which can also make the flanks important for accelerating particles \citep[see, e.g.,][]{Tylka2005,Zank2006,Hu2017,Hu2018}. As can be seen from panel (c) of Figure \ref{fig:Figure_spe30_updated}, there is a clear anisotropy of energetic particles, indicating that the energetic particles of this event are propagating predominantly outward from the Sun, which gives additional evidence that the energetic particles are likely accelerated remotely. \\

\begin{figure}[t]
	\centering
	\includegraphics[width=\textwidth]{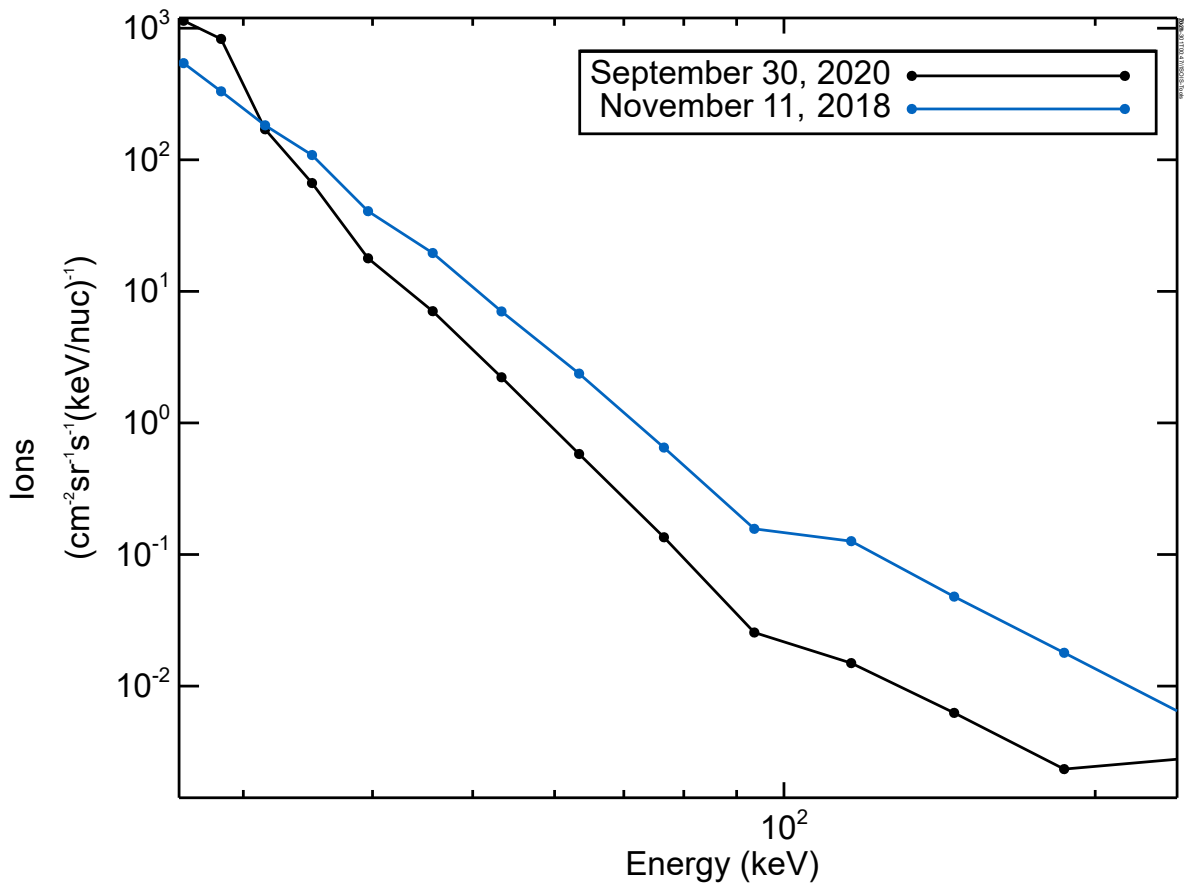}
	\caption{The event-averaged spectra of the the September 30, 2020 event (black line) and the November 11, 2018 event (blue line) plotted together for comparison. The November 11, 2018 spectrum is the same as Figure 2 (averaged over the entire day) of \citet{Giacalone2020}.}
	\label{fig:Spectra_updated}
\end{figure}

Figure \ref{fig:Spectra_updated} shows the event-averaged spectrum of this event (black line). For comparison the event averaged spectrum of the November 11, 2018 event (same as black symbols of Figure 2 of \citet{Giacalone2020}) is reproduced and shown in blue line. As can be seen in Figure \ref{fig:Spectra_updated}, the shape of the spectrum of this event and the November 11, 2018 event are somewhat similar except that the spectrum of this event is softer than the November 11, 2018 event. Note that the November 11, 2018 event is a CME-related event observed by PSP$\slash$IS{\ensuremath{\odot}}IS during its first orbit when it was about 0.25~AU from the Sun, where PSP encountered the central region of the CME. \\


\subsection{Ion Composition} \label{sec:IC}

\begin{figure}[t]
	\centering
	\includegraphics[width=\textwidth]{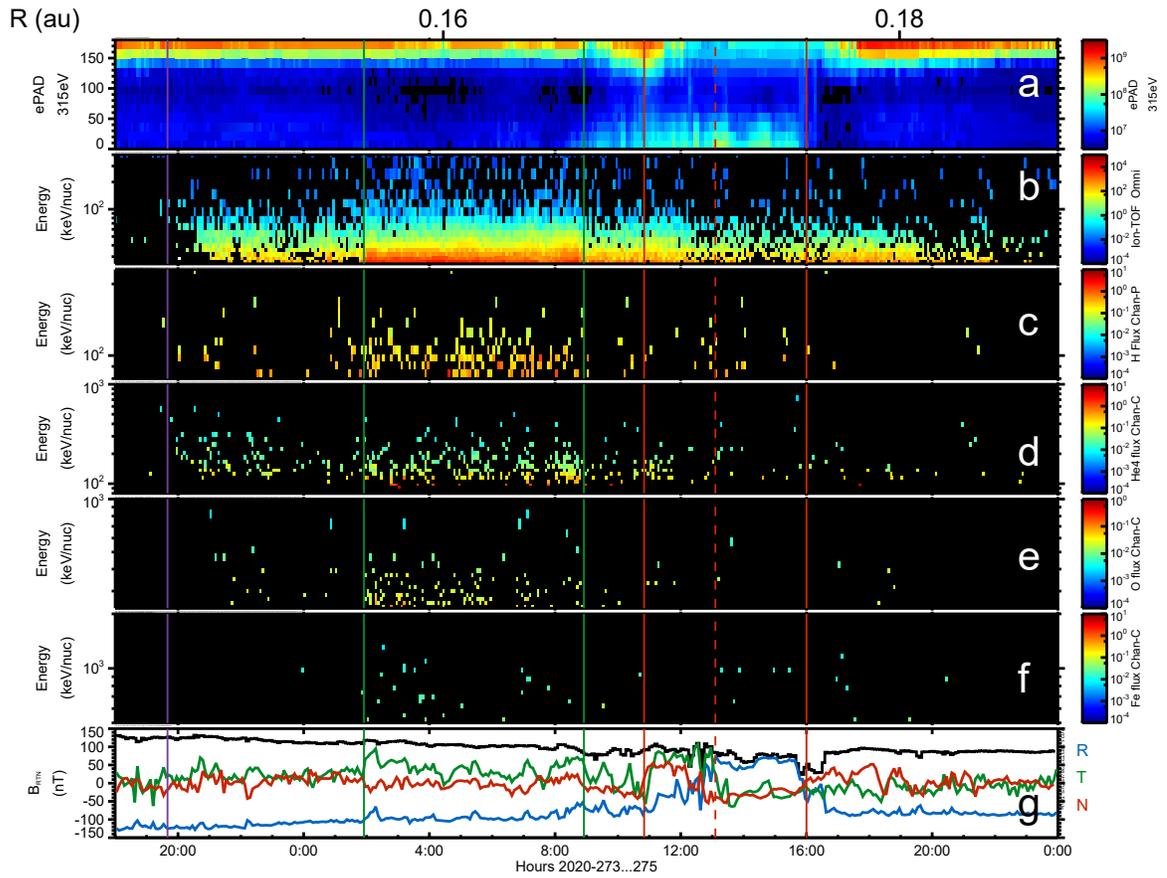}
	\caption{Ion composition of the September 30, 2020 event. The top axis shows the spacecraft’s solar radial distance in AU, same as Figure \ref{fig:Figure_spe30_updated}. Panel (a) and (b) are reproduced from Figure 6 for reference (context).  Panels (c), (d), (e), and (f) show spectrograms for proton (H), helium (4He), oxygen (O), and iron (Fe). Panel (g) is the same as panel (f) of Figure \ref{fig:Figure_spe30_updated}. All the data are averaged over 5-minute intervals. Vertical lines are described in Figures \ref{fig:Figure_spe30_updated} and \ref{fig:Figure_spe30_updated_zoomin}.}
	\label{fig:Figure_ion_composition_2_lebeled}
\end{figure}

SEP ion composition has been used as a good indicator of the origin, acceleration, and transport of SEPs \citep{Reames2021}. Figure \ref{fig:Figure_ion_composition_2_lebeled} shows the ion composition of the September 30, 2020 event. For reference, panels (a) and (b) reproduce panels (a) and (b) of Figure \ref{fig:Figure_spe30_updated}, respectively. Panels (c), (d), (e), and (f) show the spectrograms for protons, helium (4He), oxygen (O), and iron (Fe), respectively, and panel (g) reproduces panel (f) of Figure \ref{fig:Figure_spe30_updated}. As can be seen from Figure \ref{fig:Figure_ion_composition_2_lebeled}, heavy ions are observed in this event. It is populated with protons, helium, oxygen, and iron. Heavy ions are observed especially in the leading edge of the event (well before the arrival of the SBO-CME). The dispersive characteristic during the onset of this event (vertical purple line) is also clearly seen in Figure \ref{fig:Figure_ion_composition_2_lebeled}. This event has similar ion composition as the CME event of November 11, 2018 studied by \citet{Giacalone2020} and \citet{Mitchell2020a}, which was associated with an SBO-CME \citep{Korreck2020} with no shock--sheath system identified during the eruption of the CME or in the in-situ plasma data at PSP. We note that during the September 30, 2020 event, 3He and energetic electrons are not observed to be significantly above the EPI-Lo detector background, and also the FIELDS instrument did not detect radio bursts above the FIELDS instrument background. \\

\section{Discussion and Conclusions} \label{sec:discussionandconclusion}

In this paper we have studied in detail a low-energy SEP event observed by IS{\ensuremath{\odot}}IS$\slash$EPI-Lo on September 30, 2020 during the encounter 6 period inside 0.18~AU. The September 30, 2020 event is interesting in that there was no sunspot-bearing active region (at least around the longitudinal position of PSP), which gives an excellent opportunity to study the energetic particle environment of the near-quiet Sun. The event onset occurs at approximately 19:40\,UT on September 29, 2020. The event shows clear velocity dispersion during its onset, with the fastest particles arriving first. The STEREO/COR2-A coronagraph observed a CME being ejected from the solar eastern limb on September 29, 2020 around 07:00\,UT, 12~hrs before the onset of the SEP event at PSP. The CME arrived at PSP (0.17~AU) about a day after the onset of the SEP event and and PSP encountered the CME’s western flank. Throughout this SEP event, strong particle anisotropies were observed, showing that more particles are streaming outward from the Sun. \\

This small SEP event has a very interesting time profile. The energetic particle intensity rises gradually from a few to several hours and reaches a plateau later on. The strongest intensity for this event is observed in the time interval between 02:00 and 08:50\,UT on September 30, 2020, which is well before the CME passage. It is quite interesting that at the start of the plateau (first green vertical line  in Figures \ref{fig:Figure_spe30_updated}), the magnetic field and plasma properties start to change, indicating that PSP crossed a new field structure. It is possible that this new field structure is channeling energetic particles from a remote acceleration site, which most likely happened at the front of the CME. The in-situ magnetic field and strahl electron PAD indicate that these strongest energetic particle intensities are observed in the open magnetic field structure prior to the CME arrival, and there is a dropout both in intensity and number of relatively high-energy particles inside the CME structure as shown in Figures \ref{fig:Figure_spe30_updated} and \ref{fig:Figure_spe30_updated_zoomin}. As noted earlier, PSP crosses the western flank of the CME during its passage, and the time profile of this event is consistent with a western flank encounter \citep[see Figure~3.4 in ][]{Reames1999}.\\

The ion composition plot shown in Figure \ref{fig:Figure_ion_composition_2_lebeled} depicts that some heavy ions such as oxygen and iron are observed, in addition to protons and helium-4, but no significant enhancement of helium-3 and energetic electrons, giving evidence that the source of this event is unlikely to be a solar flare. Therefore, this small SEP event is not a typical small impulsive event, rather, it seems to be a small gradual SEP event. In fact, both remote-sensing (as discussed in Section \ref{sec:remote}) and in-situ (as discussed in Section \ref{sec:Sep30}) data show that this event is associated with a slow CME. Remote solar data observed using STEREO/COR2-A show that the source of this event is most likely an SBO-CME on September 29, 2020. The in-situ magnetic field and strahl electron PAD data show that the magnetic field is open and mostly radial prior to the arrival of the SBO-CME but closed upon arrival. As shown in Figures \ref{fig:Figure_spe30_updated} and \ref{fig:Figure_spe30_updated_zoomin}, throughout the event, the strahl electron flow is unidirectional except in the time interval between 11:00 and 16:00 on 274-2020 (DOY-year), where the flow is bidirectional indicating a closed magnetic field structure. This closed structure observed during the period of bidirectional electrons coincides with the SBO-CME arrival time. During its earlier orbits, PSP encountered  SBO-CMEs \citep{Korreck2020,Lario2020,Nieves-Chinchilla2020} and SEP enhancements associated with SBO-CMEs were also observed \citep{Giacalone2020,Lario2020, Mitchell2020a}, consistent with our observation. \\

We do not see a local shock in the in-situ plasma or magnetic field data throughout the event. An anisotropic and dispersive SEP event in a shockless local plasma give evidence that the dominant sources of particles were remote rather than local, consistent with earlier observations \citep[see, e.g.,][]{Giacalone2020, Mitchell2020a,Joyce2021b}. There are several candidates as a possible source of remote acceleration mechanism of these energetic particles. One possible mechanism may be that particle acceleration occurring either at plasma compressions formed in front of the propagating CME or at a weak shock initially driven by the CME that was not detected at its arrival at PSP \citep{Giacalone2020}. Another candidate as a possible source of remote acceleration mechanism of these energetic particles is the one proposed by \citet{Mitchell2020a}, which is associated with strong field-aligned current system that run in the solar atmosphere \citep{Janvier2014}. These energetic particles could be also accelerated due to magnetic-island reconnection-related processes \citep{Zank2014,Zank2015,Zhao2018,Zhao2019a,Zhao2019b}. In this scenario, the magnetic islands will be located closer to the Sun than the location where the energetic particles are observed. However, the intensity--time profile of this event gives indications that the particles are likely accelerated by a weak shock or compression driven by the CME near the Sun in agreement with earlier observations \citep{Giacalone2020}. However, it should be noted here that the SEP profile of this event is consistent with a western flank connectivity, as noted above, and the event-averaged spectrum is an indicative of a weaker event, while the event studied by \citep{Giacalone2020} was consistent with central connectivity. During this event, PSP was not radially aligned with other spacecraft and thus we could not compare this event with other energetic particle observations.\\

It is quite interesting to see that the in-situ signatures of this small gradual SEP event (associated with a weak CME) show similar properties to the well-established signatures of larger CME events, which are often associated with active regions containing sunspots, giving evidence that small SEP events may be generated by similar mechanisms as in large SEP events regardless of the size and strength of the CME. Small gradual SEP events are not totally absent at 1\,AU \citep[ see, e.g.,][]{Reames2020,Joyce2021b}, but they are rare (particularly, those with energies below 1\,MeV) due to a variety of reasons. One reason may be that they are just too small and spread out. Another possible reason may be that these are a feature of SBO-CMEs in agreement with our analysis. This event may represent direct observations of the source of low-energy SEP seed particle populations and provides a unique opportunity to show how particles evolve in the near quiet-Sun environment and also to further investigate the connection between small gradual SEP events and SBO-CMEs. This work highlights the importance of small SEP events in understanding the solar activity of the near-Sun environment that had not been previously explored. \\ 

Acknowledgements. This work was supported as a part of the Integrated Science Investigations of the Sun on NASA’s Parker Solar Probe mission, under contract no. NNN06AA01C. The IS$\odot$IS data and visualization tools are available to the community at: \url{https://spacephysics.princeton.edu/missions-instruments/isois}; data are also available via the NASA Space Physics Data Facility (\url{https://spdf.gsfc.nasa.gov/}). Parker Solar Probe was designed, built, and is now operated by the Johns Hopkins Applied Physics Laboratory as part of NASA’s Living with a Star (LWS) program (contract no. NNN06AA01C). Simulation results have been provided by the Community Coordinated Modeling Center at Goddard Space Flight Center through their public Runs on Request system (\url{http://ccmc.gsfc.nasa.gov}). The WSA model was developed by C.~N.~Arge (currently at NASA/GSFC), and the ENLIL model was developed by D.~Odstrcil (currently at GMU). We thank the STEREO team for making the SECCHI data used in this study publicly available.
E.P.’s research was supported by the NASA LWS Jack Eddy Postdoctoral Fellowship Program, administered by UCAR’s Cooperative Programs for the Advancement of Earth System Science (CPAESS) under award no. NNX16AK22G.
B.J.L. acknowledges NASA HGI 80NSSC21K0731, NASA LWS 80NSSC21K1325, and NSF AGS 1851945.


\bibliographystyle{aasjournal}



\end{document}